\documentclass[aps,twocolumn]{revtex4-2}
\usepackage{amsmath}
\usepackage[colorlinks]{hyperref}
\usepackage[utf8]{inputenc}
\usepackage{bm}
\usepackage{graphicx}

\begin{document}

\title{Signatures of gravity-mediated dark matter interaction in theories with large extra dimensions}

\author{A.~Clarke$^{a}$}
\author{V.~V.~Flambaum$^{b}$}
\author{M.~Pospelov$^{a,c,d}$}
\author{I.~B.~Samsonov$^{b}$}

\affiliation{
$^{a}$School of Physics and Astronomy, University of Minnesota, Minneapolis, MN 55455, USA \\
$^{b}$School of Physics, University of New South Wales, Sydney 2052, Australia \\
$^{c}$William I. Fine Theoretical Physics Institute, School of Physics and Astronomy, University of Minnesota, Minneapolis, MN 55455, USA\\
$^{d}$Theoretical Physics Department, CERN, 1211 Geneva, Switzerland}

\email{clar2802@umn.edu\\
v.flambaum@unsw.edu.au\\
pospelov@umn.edu\\
igor.samsonov@unsw.edu.au
}

\begin{abstract}
Dark matter particles that couple to the Standard Model only through gravity are usually regarded as inaccessible to laboratory detection. This expectation can change in theories with $n$ extra spatial dimensions, where gravity is enhanced at short distances and the potential scales as $1/r^{1+n}$.
We reconsider the gravity-mediated dark matter (DM) interactions in Arkani-Hamed-Dimopoulos-Dvali (ADD) models with $n$ large extra dimensions. The 
cumulative exchange of the gravitational Kaluza-Klein (KK) modes leads to the effective
strength of interactions with the Standard Model nucleons that scales as $m_pm_\chi M_*^{-4}$, where $ m_\chi$ is the mass of DM and $M_*$ is the fundamental $4+n$ dimensional mass scale. We confront this interaction with sensitivity achieved in the large Xe-based underground direct detection experiments and derive bounds on the $\{m_\chi,M_*\}$ parameter space that stretches all the way to $M_*\sim {\rm few}$\,TeV. We also address the indirect detection of scalar $\chi$ that can resonantly annihilate via the on-shell KK modes into the SM particles $W^\pm,Z,h$. The annihilation cross section for the process scales as $\langle\sigma v\rangle \sim m_\chi^nM_*^{-n-2}$ and stringent limits on the same parameter space can be derived from observations of high-energy galactic $\gamma$ rays. 
\end{abstract}

\maketitle

\section{Introduction}

Weakly interacting massive particles (WIMPs) remain one of the most promising candidates for dark matter. A series of experiments were set up for direct detection of this candidate using liquid noble gas \cite{XENON:2025vwd,LZ:2023poo,PandaX:2024qfu} and solid state \cite{DarkSide-50:2022qzh,SuperCDMS:2020ymb,EDELWEISS:2020fxc} DM particle detectors. So far, these experimental searches have only rendered upper limits on the WIMP-nucleon and WIMP-electron scattering cross sections, but at every increasing sensitivity, excluding or severely limiting the parameter space for many of the WIMP models. In the absence of firm model-independent theoretical predictions, the focus of the current Xe-based experimental program is to reach a sensitivity capable of detecting elastic scattering of solar and atmospheric neutrinos on nuclei. 

Among theoretical puzzles that motivate new physics at the high-energy frontier, currently explored at the LHC experiments, is the perennial question of the huge hierarchy between gravitational and weak energy scales. One radical way of addressing this question is to postulate that the scale of quantum gravity, $M_*$, is much smaller than the apparent scale that can be read off the Newton constant, $M_*\ll (G_N)^{-1/2}$. Concrete realizations of such a scenario were proposed, exploiting compact extra spatial dimensions on top of the four-dimensional Minkowski space: the Randall-Sundrum model \cite{Randall:1999ee,Randall:1999vf} predicting warped extra dimensions and the Arkani-Hamed-Dimopoulos-Dvali (ADD) model \cite{Antoniadis:1990ew,Antoniadis:1998ig,Arkani-Hamed:1998jmv,Arkani-Hamed:1998sfv} introducing large compact dimensions. 

In the ADD construction, the Standard Model (SM) particles are confined to a 3+1 dimensional ``domain wall" (also known as 3-brane), while gravity is propagating in the full $D=4+n$ space-time dimensions. Bringing $M_*$ all the way down to the TeV would solve the hierarchy problem in the ADD construction, at the expense of an extremely dense ``tower" of Kaluza-Klein states, extremely weakly coupled to the Standard Model (SM) matter. 

Gravity mediated dark matter represents an interesting option for heavy WIMP-like DM particle production and detection \cite{Lee:2013bua,Lee:2014caa,Bernal:2018qlk,deGiorgi:2021xvm,deGiorgi:2022yha}. In this scenario, the interaction between DM and Standard Model (SM) particles is mediated by massive Kaluza-Klein (KK) modes of the graviton emerging from hypothetical compact extra dimensions. An interesting corollary of the ADD construction with $M_* \propto\rm weak\,scale$ is that {\em any} stable massive particle is automatically a WIMP. Although an elementary particle is likely to satisfy the $m_\chi < M_*$ condition, a more exotic possibility can be considered, with $m_\chi > M_*$, if $\chi$ is not fundamental but composite.  In the present paper, we focus on the ADD scenario extended by one stable particle $\chi$, with the mass scale at or above the weak scale, $m_\chi \geq m_W$. We will assume that $\chi$ is SM-like, in a sense that it is restricted to the same 3+1 space-time dimensions as the SM particles.  Our goal is to explore the $\{m_\chi,M_*\}$ parameter space, and investigate whether direct dark matter detection experiments place any significant constraints. We also address the question of indirect detection due to the annihilation of $\chi-\bar\chi$ pairs to the SM states, and ensuing limits from the dark matter indirect detection program.

The models with large extra dimensions suggest that gravity can propagate equally well in both ``visible" and in extra dimensions. This leads to the enhancement of the gravitational attraction law at distances smaller than the characteristic size of the extra dimensions $R$. This short-distance force is not only a portal for DM particle interactions with SM particles, but also an additional interaction force between SM particles. Ref.~\cite{Dzuba:2022xvo} studies possible manifestations of this extra force in light atomic systems and imposes constraints on the size of extra dimensions from non-observation of anomalies in the spectra. Ref.~\cite{Flambaum:2025kkz} conjectures the existence of bound states of a large number of quarks confined by this short-distance attractive force. 

The primary goal of this paper is to contrast the WIMP dark matter in the ADD scenario with the sensitivity of the dark matter search experiments that use the dual phase TPCs (time projection chamber), mostly filled with natural xenon. These experiments have achieved breakthrough sensitivity over the course of the last twenty years. After a short overview of the basic aspects of the ADD model relevant to the content of this paper, we calculate the DM-nucleon scattering cross section due to the short-range modified gravitational potential in the ADD model, and compare it with the corresponding most stringent limits from the LZ experiment \cite{LZ:2023poo} (followed closely in sensitivity by XENONnT \cite{XENON:2025vwd} and PandaX-4T \cite{PandaX:2024qfu}).  We will show that the experimental results impose novel constraints on the $\{m_\chi,M_*\}$ parameter space, provided that $m_\chi$ is indeed at or above the electroweak scale. The main results of our work are represented by the plots of the sensitivity of the LZ detector to gravity-mediated DM particles with masses ranging from 10\,GeV to 10\,TeV; see Fig.~\ref{fig:LZlimits}. For lighter DM particles the sensitivity is significantly lower, and it cannot compete with limits on the effective $D$-dimensional Planck mass imposed by the collider searches  (see {\em e.g.}\ Ref.~\cite{ATLAS:2021kxv} for recent studies by the ATLAS collaboration).

We also consider the indirect detection limits on $\chi$ annihilation. An interesting aspect of the electroweak scale DM in ADD models is that the invariant mass of the annihilating pair is close to a large number of the KK modes so that resonant annihilation becomes possible. We evaluate the annihilation rates, finding that it can be significant for scalar DM, and is dominated by the production of $WW, ZZ$ and $hh$ final states with implications for the indirect searches of DM via observations of the galactic  high-energy gamma rays.  The corresponding limits are presented in Sec.~\ref{SectionIndirect}.

In this paper we use natural units of energy with $\hbar=c=1$.

%%%%%%%%%%%%%%%%%%%%%

\section{Overview of the ADD model for DM interaction portal}
\label{SectionOverview}

Inspired by superstring theory, the ADD model \cite{Arkani-Hamed:1998jmv} suggests that the four-dimensional Minkowski space may be extended with $n$ large compact spatial dimensions. In this model, the SM particles live only in the four-dimensional Minkowski space, while gravity is privileged to propagate equally along all $D=4+n$ dimensions. The geometry and topology of these extra dimensions is unknown, but, in the simplest scenario, it is assumed that these extra dimensions form a flat $n$-dimensional torus, and all have equal length $R$. This length may be of order micron size or smaller, based on direct tests of the Newtonian $1/r^2$ force. Limiting the size of extra dimensions in this model represents an important theoretical and experimental problem.

An immediate implication of the ADD assumption is that the $D$-dimensional Planck mass $M_*$ may be within reach of existing elementary particle accelerators such as the Large Hadron Collider (LHC). Indeed, $M_*$ is related to the conventional four-dimensional Minkowski space Planck mass $M_\text{Pl}\simeq 1.2\times 10^{19}$\,GeV through a volume factor $V_n$:
\begin{equation}
    M_\mathrm{Pl}^2 = V_n M_*^{n+2}= R^n M_*^{n+2}\,.
    \label{MD}
\end{equation}

There exist various limits both on the size of the extra dimensions $R$ and on the effective Planck mass $M_*$. The parameter $R$ is bounded from above by the Eötvös-type experiments probing Newton's law of gravity at small scales \cite{Lee:2020zjt}:
\begin{equation}
    R< 52\,\mu\text{m},
    \label{R52}
\end{equation}
which is very constraining for $M_*$ if  $n=1$ or 2.
The ATLAS collaboration puts independent limits from LHC experiments on the effective Planck mass $M_*$ from non-observation of Kaluza-Klein modes of gravitons and from the missing energy constraints \cite{ATLAS:2021kxv}. These limits for $2\leq n \leq 6$ extra dimensions are summarized in Table \ref{tab:ATLAS}. In our subsequent calculations, these limits will be assumed.

\begin{table}[h]
    \centering
    \begin{tabular}{c|c|c}
    $n$ & $M_*$ [TeV] & $R$ [m] \\\hline
    2 & 11.6 & $1.8\times 10^{-5}$ \\
    3 & 8.6 & $2.9\times 10^{-10}$\\
    4 & 7.2 & $1.1\times 10^{-12}$\\
    5 & 6.4 & $4.0\times 10^{-14}$\\
    6 & 5.9 & $4.3\times 10^{-15}$\\
    \end{tabular}
    \caption{Observed lower limits on the fundamental Planck scale $M_*$ in the ADD model from Ref.~\cite{ATLAS:2021kxv} (95\% CL). Upper bound on the size of extra dimensions $R$ is calculated using Eq.~(\ref{MD}).}
    \label{tab:ATLAS}
\end{table}

Since gravity is assumed to propagate equally well in both physical and extra dimensions, the gravitational attraction law is modified at the distances $r\ll R$: $V(r)\propto 1/r^{n+1}$. This immediately follows from the Gauss theorem for the gravitational field, since in $3+n$ spatial dimensions the spherical surface area is proportional to $r^{2+n}$. This potential may also be obtained as a sum over Kaluza-Klein modes mediating interactions between massive particles; see {\em e.g.} \cite{Giudice:1998ck}.

In this work, we will assume that the gravitational attraction between two point-like masses $m_1$ and $m_2$ is described by the following piecewise function:
\begin{equation}
    V(r) =
    \begin{cases}
        -\frac{Gm_1 m_2}{r} & \mbox{ for } r\geq R\\
        -\frac{Gm_1 m_2 R^n}{r^{n+1}} & \mbox{ for } r< R\,,
    \end{cases}
    \label{V}
\end{equation}
where $G$ is Newton's constant of gravity. Although conventional gravity is known to be too weak to play any role in interactions of elementary particles, its extra-dimensional short-range modification (\ref{V}) may be significant for heavy particles and especially for heavy DM candidates. In Ref.~\cite{Dzuba:2022xvo}, limits on the size of extra dimensions were studied from the analysis of spectra of simple atomic systems. In Ref.~\cite{Flambaum:2025kkz}, the possibility of formation of gravitationally bound states composed of a large number of quarks was explored within the ADD scenario. In the present paper, we consider the short-range modification of the gravity law (\ref{V}) as a portal for interactions of DM particles with SM matter. The important assumption made at this point is that the DM is ``SM-like", {\em i.e.} is allowed to propagate in the conventional $3+1$ dimensions. An alternative, $\chi$-particles populating extra-dimensions as well, will lead to a considerably different scenario with a multitude of KK modes for the DM itself. 
Such a scenario was discussed recently in Ref.~\cite{Garani:2025wuu}. 

The short-range modification of the gravitational interaction (\ref{V}) can actually be extended to large distances because the potential $V(r)\propto 1/r^{n+1}$ is negligible at $r\gg R$. Therefore, within our goal of accuracy, in elementary particles interactions we can safely replace the potential (\ref{V}) with 
\begin{equation}
    V(r) = -\frac\beta{r^{n+1}}\,,\qquad
    \beta = G m_1 m_2 R^n = \frac{m_1 m_2}{M_*^{n+2}}\,.
    \label{Vr}
\end{equation}
This potential is singular for $n\geq2$, but is still relatively long-range for $n=1$. Therefore, we will consider these cases separately below.

The singularity of the potential (\ref{Vr}) at $r\to 0$ is unphysical, and it must be regularized using a short-range cutoff parameter $r_c\ll R$. In this paper, we consider the following short-distance regularization of this potential:
\begin{equation}
  V_\text{reg}(r) = -\frac{\beta}{(r^2 + r_c^2)^{(n+1)/2}}\,,
  \label{Vreg}
\end{equation}
although other regularizations are also possible. It is natural to identify the cutoff parameter $r_c$ with the effective Planck length in $D=4+n$ spacetime,
\begin{equation}
    r_c = L_* = \frac1{M_*}\,.
    \label{rc}
\end{equation}
Indeed, at the distances $r\ll L_*$ the classical description of gravity in terms of the potential (\ref{Vr}) fails since (unknown) quantum gravity effects should start to play a significant role. More accurately, $r_c$ should be related to $M_*$ up to some unknown numerical factor $C$, $r_c = CM_*^{-1}$. In some cases, such as in the $n=2$ case, the precise value of $C$ does not affect the strength of the effective SM-SM and SM-DM interactions mediated by the KK exchange too much, as we will see below. In other cases, such as $n\geq 3$ there is a power-like dependence on $C$, and since we are taking $C=1$ our results are more in the spirit of a parametric estimate, rather than a precise result. (If a realistic ADD-style particle physics model is indeed constructed and embedded into the string theory, then $C$ might be calculable from first principles.)

Note that in Ref.~\cite{Flambaum:2025kkz} the cutoff parameter was chosen to be equal to the gravitational (higher-dimensional Schwarzschild) radius $r_S$ for composite objects. For the interactions of elementary particles considered in this paper, it is possible to show that $r_S\ll L_*$, and the choice (\ref{rc}) is justified.

%%%%%%%%%%%%%%%%%%%%%%%%%%%%%%%%%%%%%%%%%%%%%%%%%%%

\section{Sensitivity of liquid noble gas detectors to DM within ADD scenario}
\label{SectionSensitivity}

In this section, we consider a DM particle candidate with mass $m_\chi> 10$\,GeV and assume that it may interact with SM particles via the potential (\ref{Vr}), or rather its regularized version (\ref{Vreg}). We will consider scattering of these particles off nucleons and nuclei with the goal of sensitivity estimates comparable to those of XENONnT \cite{XENON:2025vwd} and LZ \cite{LZ:2024zvo}. First we consider the case $n\geq2$ since the corresponding potential (\ref{Vr}) is highly singular, and separately study $n=1$ which is relatively long range. These calculations can be conducted using standard diagrammatic techniques using the summation over masses of KK modes with an effective cutoff, or directly in the position space using $V(r)$. 

\subsection{DM-nucleon scattering cross section for $n\geq2$} 

In the case $n\geq2$, the potential (\ref{Vreg}) is singular and dominates at very short distances, which are much smaller than the size of the nucleus. The effect of this interaction is undistinguishable from contact DM-nucleus scattering. Therefore, we simplify the problem by replacing the potential (\ref{Vreg}) with an equivalent contact interaction,
\begin{equation}
    V_\text{reg}(r) \to V_\text{contact}(r) = V_n \delta^3(r)\,,
\end{equation}
where the interaction constant $V_n$ is found from the condition 
\begin{equation}
\int d^3 r\, V_\text{reg}(r) = \int d^3r\, V_\text{contact}(r) = V_n\,.
\end{equation}
Note that for $n\geq3$ this integration can be safely performed over the full three-dimensional space, as the integral is dominated by short distances. The result scales as $r_c^{2-n}\propto (C/M_*)^{2-n}$ and therefore depends on the details of a chosen regularization.

The case $n=2$ is special due to a logarithmic divergence. The experiments operate in a certain window of energy recoil $E_R$ that defines a characteristic momentum transfer $q\sim \sqrt{2E_R m_{\rm Xe}}$. To logarithmic accuracy, the upper limit of $d^3r$ integration can be identified with $q^{-1}$. This follows from the direct calculation of the Born amplitude. The results of the integration are:
\begin{align}
    V_n =& -\frac{\pi^{3/2}m_1m_2}{M_*^4} \frac{ \Gamma(-1+n/2)}{\Gamma((n+1)/2)}\,, \qquad (n\geq3)
    \label{Vn}
     \end{align}
    \begin{align}
    V_2 = -\frac{4\pi m_1 m_2}{M_*^4}\ln(M_*/q) ,\, \qquad (n=2)
    \label{V2}
\end{align}
where $r_c\ll R$ is assumed in the last approximation. For the $n=2$ case the results are numerically robust, as $E_R \sim O(10\,\rm keV)$ leading to $q\sim 50$\,MeV and $\ln(M_*/q)\sim 10$, taking $M_*$ in the TeV range.  

In general, it is possible to calculate the scattering cross section of DM particles off heavy nuclei with the potential (\ref{Vreg}). This problem drastically simplifies in the case of the contact interaction. It is sufficient to consider the scattering of the DM particle off an isolated nucleon with mass $m_p$ because experimental collaborations usually present the limits on the DM-nucleon scattering cross section assuming small dependence of the latter on the momentum transfer; see, e.g., Refs.~\cite{XENON:2025vwd,LZ:2024zvo}. Notice that the influence of the nuclear form factor is taken into account by the experimental collaborations. For non-relativistic particles with a contact interaction, this scattering cross section has the standard form:
\begin{equation}
    \sigma_0 = \frac{\mu_p^2 V_n^2}{\pi}\,,
    \label{sigma0}
\end{equation}
where $\mu_p = \frac{m_p m_\chi}{m_p + m_\chi}$ is the reduced mass. 

An interesting question is whether $m_\chi > M_*$ is allowed. The standard lore is that an elementary point-like particle with a mass larger than the quantum gravity mass scale will become a black hole. On the other hand, for our treatment of direct detection limits it is not important whether $\chi$ is elementary or composite, and therefore we choose to extend our treatment into the range of trans-Planckian (trans-$M_*$) masses. A somewhat limiting factor here is the spatial extent of the $\chi$ particles if they are composite objects. For our treatment, it does not really change our calculations as long as the physical size $R_\chi$ is below $q^{-1}$. Extremely large composite objects might have their own ``dark form factors"; see {\em e.g.} Ref.\,\cite{Wise:2014jva}. 

\begin{figure}
    \centering
    \includegraphics[width=1\linewidth]{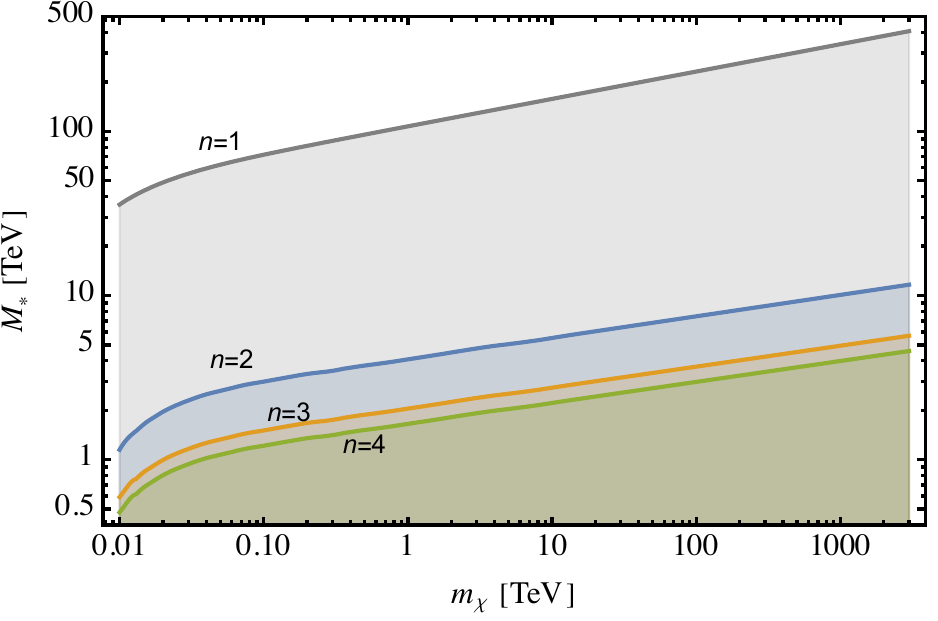}
    \caption{Lower limits on the effective Planck mass $M_*$ in the ADD model with $n=1,2,3$ and 4 extra dimensions. These lower bounds are found from the null results of the LUX-ZEPLIN experiment \cite{LZ:2024zvo} in searches of the DM particles with short-range DM-nucleus interaction. These plots represent sensitivity of the LZ experiment to DM particles with gravitational interaction portal within the ADD model.
    }
    \label{fig:LZlimits}
\end{figure}

The strongest upper bounds on this scattering cross section are found in Ref.~\cite{LZ:2024zvo} for DM particles in the mass range $10\,\text{GeV}<m_\chi <10\,\text{TeV}$. Using these bounds, we derive the lower limits on the Planck mass $M_*$ for $n=2,3,4,\ldots$ extra dimensions; see Fig.~\ref{fig:LZlimits}. The limits in the region $10\,\text{TeV}<m_\chi<1\,\text{PeV}$ are found by the corresponding extrapolation on the experimental data from Ref.~\cite{LZ:2024zvo}.

\begin{table}[t]
    \centering
    \begin{tabular}{c|c|c}
 $n$ & $M_*$ [TeV] & $R$ [m] \\\hline
 2 & 5.5 & $7.9\times 10^{-5}$ \\
 3 & 2.7 & $1.9\times 10^{-9}$ \\
 4 & 2.2 & $6.6\times 10^{-12}$ \\
 5 & 1.9 & $2.1\times 10^{-13}$ \\
 6 & 1.8 & $2.1\times 10^{-14}$ 
    \end{tabular}
    \caption{Lower limits on effective Planck mass $M_*$ and the corresponding size of the extra dimensions $R$ in the ADD model from non-observation of DM particles with mass $m_\chi=10$\,TeV at LZ detector.}
    \label{tab:MRlimits}
\end{table}

Fig.~\ref{fig:LZlimits} shows that the liquid noble gas detectors of DM particles possess the highest sensitivity to detect heavy DM particles in the case of $n=2$ extra spatial dimensions. Notably, for $n=2$, the sensitivity of the LZ experiment exceeds that reported by the ATLAS collaboration \cite{ATLAS:2021kxv} if the DM particle mass is $m_\chi\gtrsim 0.5$\,PeV. However, for $n\geq3$, the sensitivity to such DM particle interaction is significantly lower. Therefore, in Fig.~\ref{fig:LZlimits}, we present only the sensitivity curves for $n=1,2,3$ and 4, understanding that higher cases have a lower chance of potential detection. 

Lower limits on the effective Planck mass $M_*$ and the corresponding size of the extra dimensions $R$ in the ADD model from non-observation of DM particles with mass $m_\chi=10$\,TeV at LZ detector are given in Table~\ref{tab:MRlimits}.

\subsection{DM-nucleus scattering cross section in case of one extra dimension}

The modified gravitational potential (\ref{Vr}) for two point-like masses $m_\chi$ and $m_N$ is
\begin{equation}
    V(r) = - \frac{\beta}{r^2}\,,\qquad \beta = G m_\chi m_N R = \frac{m_\chi m_N}{M_*^3}\,.
\end{equation}
The corresponding Fourier transform is
\begin{equation}
    \tilde V(q) = \int V(r) e^{i\vec q \cdot\vec r} d^3r = -2\pi^2\frac{\beta}{q}\,.
\end{equation}

The differential scattering cross section off an extended nucleus with form factor $F(q)$ in the first Born approximation is
\begin{equation}
    \frac{d\sigma}{d\Omega} = \frac{\mu_N^2}{4\pi^2}\left[\tilde V(q)F(q) \right]^2 = \pi^2 \beta^2\mu_N^2 \frac{F^2(q)}{q^2} \,.
\end{equation}
Recalling that the nuclear recoil energy is $E_\mathrm{rec} = \frac{q^2}{2m_N}$, we find the scattering cross section per unit nuclear recoil energy:
\begin{equation}
    \frac{d\sigma}{dE_\mathrm{rec}} = \frac{\pi^3\beta^2 }{E_\mathrm{rec}} \frac{1}{ v^2}F^2(q) = F^2(q) \frac{\pi^3 m_N^2 m_\chi^2}{v^2 E_\mathrm{rec} M_*^6}\,,
    \label{dSigmadE}
\end{equation}
where $v$ is the speed of the DM particle in the laboratory frame. Multiplying Eq.~(\ref{dSigmadE}) by the DM particle flux and averaging it with the DM speed distribution $f(v)$ (normalized as $\int f(v)v\,dv = 1$) we find the differential recoil rate
\begin{align}
    \frac{dR_\text{rec}}{dE_\mathrm{rec}} =& \frac1{m_N}\frac{\rho_\chi}{m_\chi} \int_{v>v_\text{min}} dv \,v\,f(v) \frac{d\sigma}{dE_\mathrm{rec}} 
    \nonumber\\=&
    \frac{\pi^3 \rho_\chi}{m_N m_\chi}\frac{m_N^2 m_\chi^2}{M_*^6} \frac{F^2(q)}{E_\mathrm{rec}} \int_{v>v_\text{min}} dv \frac{f(v)}{v} 
    \label{dRdE}
\end{align}
where $\rho_\chi\simeq 0.4\,\text{GeV/cm}^3$ is the local dark matter density and
\begin{equation}
    v_\text{min} = \sqrt{\frac{m_N E_\mathrm{rec}}{2\mu_N^2}}\,.
\end{equation}
With the standard dark matter halo model (see, e.g., \cite{Lewin:1995rx}), the DM speed distribution may be chosen in the form of a Gaussian function
\begin{equation}
    f(v) = \frac{v}{\sqrt{\pi}v_0 v_\text{obs}} e^{-(v+v_\text{obs})^2/v_0^2}(e^{4v v_\text{obs}/v_0^2}-1)\,,
    \label{f(v)}
\end{equation}
with $v_0=220$\,km/s, $v_\text{ob} = 232$\,km/s. With this distribution, the integral over the DM speed is calculated analytically:
\begin{align}
    g(E_\mathrm{rec}) \equiv& \int_{v>v_\text{min}} dv \frac{f(v)}{v} \\ =& \frac1{2v_\text{obs}}\left[ \text{erf}\left(\frac{v_\text{min} + v_\text{obs}}{v_0}\right) - \text{erf}\left(\frac{v_\text{min}-v_\text{obs}}{v_0}\right) \right].\nonumber
\end{align}

Finally, integrating Eq.~(\ref{dRdE}) from the detector threshold $E_\text{min} =1$\,keV to $E_\text{max} = \frac{2\mu_N^2 v_\text{max}^2}{m_N}$, ($v_\text{max} \simeq 600$\,km/s is the galactic escape velocity) we find
\begin{equation}
    R_\text{rec} = \int_{E_\text{min}}^{E_\text{max}} \frac{dR_\text{rec}}{dE_\mathrm{rec}}dE_\mathrm{rec} = I_0 \frac{\rho_\chi m_\chi m_N}{M_*^6}
    \label{Rrec}
\end{equation}
where
\begin{equation}
    I_0 = \pi^3 \int_{E_\text{min}}^{E_\text{max}} \frac{F^2(\sqrt{2m_N E_\text{rec}})}{E_\text{rec}}g(E_\text{rec}) dE_\text{rec}\,.
    \label{I0}
\end{equation}

Note that $I_0$ in Eq.~(\ref{I0}) depends both on the DM mass $m_\chi$ and nuclear mass $m_N$. Using the solid sphere nuclear form factor
\begin{equation}
    F(q) = 3\frac{\sin(qR_N) - qR_N \cos(qR_N)}{(qR_N)^3}\,,
\end{equation}
with nuclear radius $R_N$, we calculate it numerically for the xenon nucleus with mass $m_N \simeq 131 m_p$. We find that this function $I_0$ changes monotonically from $1.1\times 10^4$ to $8.3\times 10^4$ for DM particles with masses ranging from 10\,GeV to 10\,TeV.

The experimental limit on the nuclear recoil rate set by the LZ collaboration \cite{LZ:2024zvo} is
\begin{equation}
    R_\text{rec,exp} = \frac{2.3}{4.2\text{ tonne}\times\text{year}}.
\end{equation}
Comparing this value with the estimated recoil rate (\ref{Rrec}) we find the limit on the effective Planck mass shown in Fig.~\ref{fig:LZlimits}.

\subsection{Comments on the Migdal (inelastic) scattering}

The Migdal effect has been widely used in searches for DM particles in direct detection experiments with liquid noble gases; see, e.g., Refs.~\cite{Bernabei:2007jz,Ibe:2017yqa,Dolan:2017xbu}. Theoretically, it allows one to detect DM particles with sub-GeV masses in experiments such as XENONnT \cite{XENON:2025vwd} and LUX-ZEPLIN \cite{LZ:2024zvo}. Experimentally, this effect was probed in the neutron scattering experiments \cite{Yi:2026fmf,Xu:2023wev}. In a nutshell, the Migdal effect represents inelastic DM scattering with negligible nuclear recoil but with a non-zero probability of atomic ionization. In this subsection, we discuss the sensitivity of the DM direct detection experiments to sub-GeV DM particles with the gravitational interaction portal in the ADD model.

\begin{figure}
    \centering
    \includegraphics[width=\linewidth]{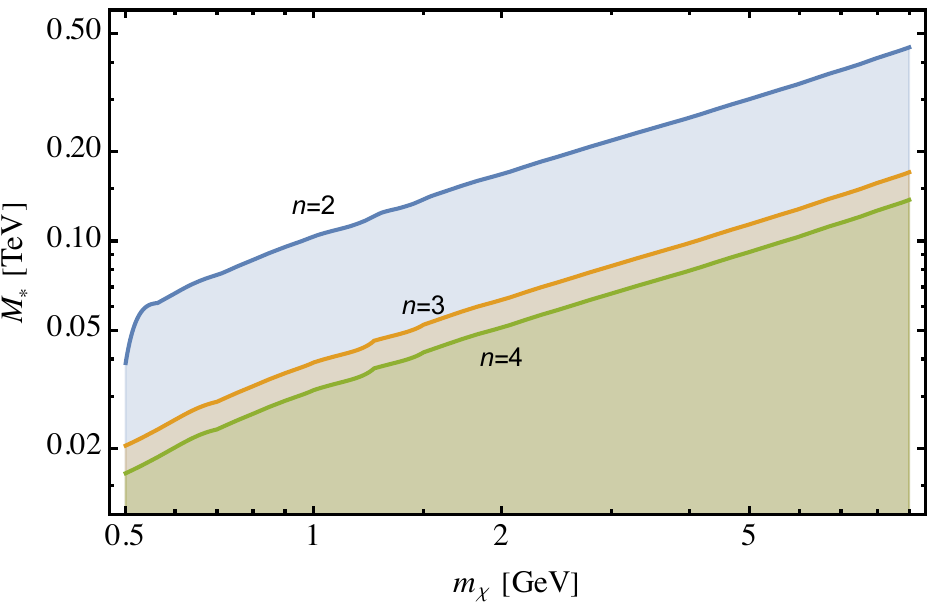}
    \caption{Sensitivity of LZ liquid noble gas detectors to sub-GeV DM particles with gravitational interaction portal within the ADD model. Different curves correspond to $n=2,3$ and 4 extra spatial dimensions. These limits are obtained due to the Migdal effect \cite{Bernabei:2007jz,Ibe:2017yqa,Dolan:2017xbu}.}
    \label{fig:Migdal}
\end{figure}

The theoretical basis for the detection of DM particles by using the Migdal effect was thoroughly worked out in Refs.~\cite{Bernabei:2007jz,Ibe:2017yqa,Dolan:2017xbu}. Here we do not need to repeat these calculations because the experimental collaborations usually present limits on the DM-nucleon scattering cross section at zero momentum transfer taking in account the Migdal effect; see Ref.~\cite{LZ:2023poo}. In our model, the same cross section is calculated in Eqs.~(\ref{Vn},\ref{V2}) and (\ref{sigma0}). This allows us to immediately find the limits on the effective Planck mass $M_*$ for DM particles with masses ranging from 0.5\,GeV to 9\,GeV; see Fig.~\ref{fig:Migdal}. As is seen from these plots, for the detection of such sub-GeV DM particles the effective Planck mass should be well below 1\,TeV scale. This region is strongly excluded by the ATLAS collaboration \cite{ATLAS:2021kxv}, see Table~\ref{tab:ATLAS}. Thus, the Migdal effect can hardly help in the detection of sub-GeV DM particles within this scenario. The main reason, of course, is due to the suppression of the gravitational interaction by $m_\chi/M_*$, which is very disadvantageous for lighter DM. 

%%%%%%%%%%%%%%%%%%%%%%%%%%%%%%%%%%%%%%%%%%%%%%%%%%%
\section{Indirect Detection Sensitivity}
\label{SectionIndirect}

In this section, we consider indirect detection constraints, which provide a complementary probe to the direct detection limits derived earlier. Typically, gravity-mediated annihilation of DM is highly suppressed. However, if the center-of-mass energy is sufficiently close to a KK mode mass, then DM annihilation may proceed through a KK resonance, significantly enhancing the annihilation rate. Motivated by this possibility, we consider the annihilation of a scalar DM candidate into a spin-0 KK resonance, which subsequently decays into SM particles. We neglect spin-2 KK modes because they are subdominant due to the $d$-wave suppression of annihilation. The annihilation of fermionic dark matter will proceed in the $p$ wave, and will be suppressed as well. 

Since we consider energy scales much larger than the electro-weak scale, we take the decay into SM particles as being dominated by the 4 degrees of freedom of the Higgs doublet $H$, 
\begin{equation}
    \chi^\dagger\chi \to \tilde \phi_{\vec n} \to H^\dagger H\,.
\end{equation} 
The Higgs doublet decay channel dominates this process as the KK scalar couples most efficiently with scalars. The rest of the SM fields have a conformal coupling to gravity in the limit of $m_W \ll M_*$ and their contributions to the final states are small.  We parameterize the interaction of the spin-0 KK mode $\tilde \phi_{\vec n}$ with the scalar field $H$ using the interaction Lagrangian given in \cite{Han:1998sg},
\begin{equation}
    \mathcal{L}_{\rm int} = \omega \kappa\tilde\phi_{\vec n}(D^{\mu}H^{\dagger} D_{\mu}H - 2m_{H}^{2}H^{\dagger}H)\,,
\end{equation} where $\kappa = \sqrt{16 \pi G}$ and $\omega = \sqrt{\frac{2}{3(n+2)}}$. Because galactic DM is highly non-relativistic, the cross-section for this process is well approximated by the non-relativistic Breit-Wigner formula, 
\begin{equation}
    \sigma_{\vec n} = \frac{4\pi}{\vec k^{2}}\frac{\Gamma_\text{DM}\Gamma_\text{SM}/4}{(\sqrt{s} - m_{\vec n})^{2}+\Gamma^{2}/4}\,,
\end{equation}
where $\vec k^2$ is the squared three-momentum of either incoming particle in the center-of-mass frame, $\Gamma_\text{DM}$ is the width of the KK mode decay into the dark-matter pair, $\Gamma_\text{SM}$ is the width of the KK mode decay into Standard Model particles $H$, and $\Gamma = \Gamma_\text{DM} + \Gamma_\text{SM}$. 

It is possible to show that the total decay width is much smaller than the KK mode spacing for $n\leq 4$. Thus, we can use the narrow-width approximation, yielding 
\begin{equation}
    \sigma_{\vec n} = \frac{8\pi^{2}}{m_{\chi}^{2}v^{2}}\frac{\Gamma_\text{DM}\Gamma_\text{SM}}{\Gamma}\delta(\sqrt{s} - m_{\vec{n}})\,,
\end{equation}
where $v$ is the relative DM particle velocity. This equation further simplifies near the resonance, where $\Gamma_\text{DM} \ll \Gamma_\text{SM}$,
\begin{equation}
    \sigma_{\vec n}  = \frac{8 \pi^{2}}{m_{\chi}^{2}v^{2}}\Gamma_\text{DM}\delta(\sqrt{s} -m_{\vec n})\,.
\end{equation}
Summing over KK modes and averaging over the velocity distribution $f(v)$, then yields 
\begin{equation}
    \langle \sigma v\rangle = \int dv\int dm_{\vec n}^{2}\sigma_{\vec n}vf(v)\rho(m_{\vec n})\,.
    \label{sigmav}
\end{equation}

The density of KK states was found in Ref.~\cite{Han:1998sg}:
\begin{equation}
    \rho(m_{\vec n}) = \frac{ M_\mathrm{Pl}^2m_{\vec{n}}^{n-2}}{(4\pi)^{n/2}\Gamma(n/2)M_*^{n+2}}\,.
    \label{densityStates}
\end{equation} 
Note also that the decay width of a scalar $\tilde\phi_{\vec n}$ into two scalars $\chi^\dagger\chi$ is
\begin{equation}
    \Gamma_\text{DM} = \frac{\omega^2 \kappa^2 m_\chi^3 v}{64\pi}\,.
    \label{GammaDM}
\end{equation}
Using Eqs.~(\ref{densityStates},\ref{GammaDM}) and the normalization of the velocity distribution, $\int f(v)dv = 1$, we perform the integration in Eq.~(\ref{sigmav}):
\begin{equation}
    \langle \sigma v\rangle = \frac{\alpha_{n}m_{\chi}^{n}}{M_*^{n+2}}\,,
\end{equation}
where $\alpha_{n}$ are dimensionless coefficients
\begin{equation}
    \alpha_n = \frac{4\pi^{2-n/2}}{3(n+2)\Gamma(n/2)}\,.
\end{equation}
For the $n=2$ case, this gives the prediction $\langle \sigma v\rangle =(\pi/3)\times m_\chi^2/M_*^4$.

\begin{figure}
    \centering
    \includegraphics[width=\linewidth]{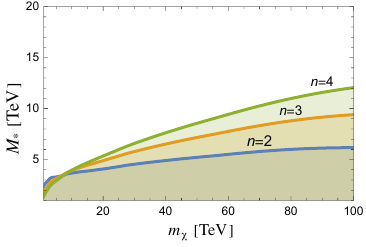}
    \caption{Lower limits on $M_*$ for $n = 2$, 3, and 4 extra dimensions resulting from limits on $\langle \sigma v\rangle$ given by indirect detection experiments \cite{MAGIC:2016xys}.}
    \label{fig:indirectDetection}
\end{figure}

As is easily seen, resonant annihilation provides a significant parametric enhancement in comparison to the standard non-resonant case, which is suppressed by additional powers of $M_*$ and typically falls below the sensitivity of indirect detection experiments. Resonant annihilation therefore provides a channel that can be used to provide stringent constraints on gravitational annihilation. Comparing our result for $\langle\sigma v\rangle$ with the upper limits on $\langle \sigma v\rangle$ given in Ref.~\cite{MAGIC:2016xys}, we derive the lower limits on $M_*$ for $n= 2$, 3 and 4 additional dimensions given in Fig.~\ref{fig:indirectDetection}.  Notice that unlike the direct detection case, these constraints apply to elementary DM particles. In that sense, one should expect $m_\chi < M_*$, which would apply only to sub-10\,TeV range in Fig.~\ref{fig:indirectDetection}.
%%%%%%%%%%%%%%%%%%%%%%%%%%%%%%%%%%%%%%%%%%%%%%%%%%%

\section{Conclusions}

In this paper, we have considered the current experimental sensitivity to WIMP dark matter within ADD scenarios with large extra dimensions. 
Typical WIMP models are built postulating some type of non-gravitational interaction between dark matter particles and the SM. In contrast, in ADD construction any stable particle with mass at and above the EW scale is a WIMP. This follows from a very generic consequence that a cumulative effect of the gravitational interaction between any two massive particles leads to a contact interaction suppressed by the scale of quantum gravity ($r_c$ in the position space or $M_*$ in the momentum space). Although we remained totally agnostic about the origin of DM particles, some studies suggest that $\chi$ can naturally emerge as ``bending modes" of domain walls; see, {\em e.g.} \cite{Cembranos:2003mr}. 

Direct detection experiments, among which the LZ is the current leader, provide non-trivial sensitivity to the ADD scenario extended by a stable particle $\chi$ that we assume to be elementary and saturate the DM abundance. Summarizing the parametric dependence of scattering cross sections for heavy $m_\chi$ one finds for ($n\geq 2$) $\sigma_{\chi p} \propto m^2_\chi m_p^4 M_*^{-8}$. For $n=2$, the numerical coefficient in front can be calculated without significant uncertainties, but for higher $n$ it depends quite sensitively on the cutoff scale. Due to the two extra powers of $m_p$ in this formula, the cross section is much suppressed compared to the typical $\sigma_{\chi p}$ mediated by the weak force. Nevertheless, current Xe-based experimental searches are so sensitive, that even that small cross section is meaningfully constrained, Fig.~\ref{fig:LZlimits}. 

With an additional assumption of a particle-antiparticle symmetric DM sector, one can calculate the annihilation, which is enhanced by resonant production of KK modes, and for the scalar DM can go in the $s$-wave. The resonant annihilation process is cutoff independent, and is found to scale as $\langle\sigma v\rangle \propto m_\chi^n M_*^{-n-2}$. When $m_\chi$ is parametrically heavier than the weak scale, then the annihilation to the SM is dominated by particles from the Higgs multiplet, specifically longitudinal $W^+W^-$, $ZZ$ and the physical Higgs pairs. All of these intermediate states fragment to produce $\gamma$-rays and are therefore subject to indirect detection constraints. 

Of course, any WIMP model is incomplete without a concrete scenario for achieving the correct cosmological abundance. The ``standard" WIMP abundance is achieved via the freeze-out of annihilation at $T_\text{f.o.}\sim 0.05 m_\chi$. For the ADD construction such a scenario would  not work because the starting point for a thermal Universe has to be very low, in the tens of MeV, known as ``normalcy temperature" \cite{Arkani-Hamed:1998sfv,Arkani-Hamed:1999fet}. The requirement of low reheating is dictated by otherwise overproduced KK modes. Such a low reheating temperature could be achieved if the decay of the inflaton is very slow. However, it does not preclude access to higher energy scales, provided that its mass is large (see {\em e.g.} Ref.~\cite{Allahverdi:2003aq}), so that non-thermal pair-production of $\chi$'s is possible. 

\vspace{3mm}
\textit{Acknowledgments.} -- The work of IBS and VVF was supported by the Australian Research Council Grant No.\ DP230101058. The work of MP is supported in part by the DOE grant DE-SC0011842.

%%%%%%%%%%%%%%%%%%%%%%%%%%%%%%%%%%%%%%%%%%%%%%%%%%%
%

\end{document}